# Probability distribution and entropy as a measure of uncertainty

Qiuping A. Wang

Institut Supérieur des Matériaux et Mécaniques Avancés du Mans, 44 Av. Bartholdi,

72000 Le Mans, France

**Abstract**

The relationship between three probability distributions and their maximizable entropy forms is discussed without postulating entropy property. For this purpose, the entropy $I$ is defined as a measure of uncertainty of the probability distribution of a random variable $x$ by a variational relationship $dI = d\overline{x} - \overline{dx}$, a definition underlying the maximization of entropy for corresponding distribution.

## 1) Introduction

It is well known that entropy and information can be considered as measures of uncertainty of probability distribution. However, the functional relationship between entropy and the associated probability distribution has since long been a question in statistical and informational science. There are many relationships established on the basis of the properties of entropy. In the conventional information theory and some of its extensions, these properties are postulated, such as the additivity and the extensivity in the Shannon information theory. The reader can refer to the references [1] to [8] to see several examples of entropies proposed on the basis of postulated entropy properties. Among all these entropies, the most famous one is the Shannon informational entropy ($S=-\sum_i p_i \ln p_i$)[2] which was almost the only one widely used in equilibrium thermodynamics (Boltzmann-Gibbs entropy) and in nonequilibrium dynamics (Kolmogorov-Sinai entropy for example). But the question remains open in the scientific community about whether or not Shannon entropy is the unique useful measure of statistical uncertainty or information[9].

The origin of this question can be traced back to the principle of maximum entropy (maxent) by Jaynes[10] who claimed the Shannon entropy was singled out to be the only consistent measure of uncertainty to be maximized in maxent. In view of the fact that this uniqueness was argued from the Shannon postulates of information property[2], one naturally asks what happens if some of these properties are changed. Some of the entropies in the list of [1] are found by mathematical considerations which change the logic of Shannon. Recently, a nonextensive statistics[6][7][8] (NES) proposed to use some entropies for thermodynamics and stochastic dynamics of certain nonextensive systems. NES has given rise to a large number of papers with very different viewpoints dealing with equilibrium and nonequilibrium systems and incited considerable debate[11][12][13] within the statistical physics community. Some key questions in the debate are: whether or not it is necessary to replace Boltzmann-Gibbs-Shannon entropy with other ones in different physical situation? what are the possible entropy forms which can be maximized in order to derive probability distribution according to maxent?

One remembers that in the actual applications of maxent the used entropy forms are either directly postulated or derived from postulated properties of entropy[1-8]. The correctness of these entropies is then verified through the validity of derived probability distributions. In the present work, we will invert this reasoning in order to find maximizable entropy form directly from known probability distributions without postulating the properties of entropy. For this purpose, we need a generic entropy definition underlying in addition variational approach or maxent. Inspired by the first and second laws of thermodynamics for reversible process, we introduce a variational definition of entropy *I* such as $dI=d\overline{x}-\overline{dx}$ for the measure of probabilistic uncertainty of the simple situation with only one random variable *x*. We stress that the main objective of this work is to show the non-uniqueness of Shannon entropy as maximizable uncertainty measure. Other entropy forms must be introduced for different probability distributions. We would like to stress also that this is a conceptual work tackling the mathematical form of entropy without considering the detailed physics behind the distribution laws used in the calculations. In what follows, we first talk about three probability distributions and their invariant properties. The maximizable entropy form for each of them is then derived thanks to the definition $dI=d\overline{x}-\overline{dx}$.

## 2) Three probability laws and their invariance

In this section, by some trivial calculations one can find in textbooks, we want to underline the fact that a probability distribution may be derived uniquely from its invariance. By invariance of a function *f(x)*, we means that the dependence on *x* is invariant at transformation of *x* into *x'*, i.e., $f(x')\propto f(x)$. We consider three invariances corresponding to exponential, power law and q-exponential distributions, respectively.

### a) *Translation invariance and exponential law*

Suppose that $f(x)$ is invariant by a translation of $x \to x+b$, i.e.

$$f(x+b)=g(b)f(x) \tag{1}$$

where $g(b)$ depends on the form of *f(x)*. We have $\frac{df(x+b)}{db}=\frac{df(x+b)}{d(x+b)}=g'(b)f(x)$ and $\frac{df(x)}{dx}=g'(0)f(x)$ or $\frac{df(x)}{f}=g'(0)dx$ (*b*=0). This means

$$\ln f(x)=g'(0)x+c \text{ or } f(x)=ce^{g'(0)x}. \tag{2}$$

where *c* is some constant. If *f(x)* gives a probability such as $p(x)=\frac{1}{Z}f(x)$ where $Z=\sum_x f(x)$, the normalization condition $\sum_x p(x)=1$ will make *p(x)* strictly invariant versus the transformation $x \to x'=x+b$, i.e., $p(x')=\frac{1}{Z'}f(x')=\frac{1}{Z}f(x)=p(x)$ since $Z'=\sum_{x'}f(x')=\sum_x f(x+b)=g(b)\sum_x f(x)=Zg(b)$.

## b) Scale invariance and power law

Now suppose that $f(x)$ is scale invariant, we should have

$$f(bx) = g(b)f(x) \tag{3}$$

where $b$ is the scale factor of the transformation. We make following calculation $\frac{df(bx)}{db} = \frac{df(bx)}{d(bx)}x = g'(b)f(x)$ to get $\frac{df(x)}{dx}x = g'(1)f(x)$, which means

$$f(x) = c\, x^{g'(1)}. \tag{4}$$

This kind of laws is widely observed in nature for different dynamical systems such as language systems[14] and scale free networks[15] among many others[16]. The well known Levy flight for large $x$ is a good example of power law with $g'(1)=-1-\alpha$ where $0<\alpha<2$.

## c) The q-exponential and its invariance properties

Here we would like to mention a probability which has attracted a lot of attention in the last years:

$$f(x)=c[1+a\beta x]^{\frac{1}{a}}. \tag{5}$$

where $a$ and $\beta$ are some constants. The Zipf-Mandelbrot law $f(x) = c[1+x]^{-\alpha}$ observed in textual systems and other evolutionary systems[17] can be considered as a kind of q-exponential law. Another example of this law is the equilibrium thermodynamic distribution for finite systems in equilibrium with a finite heat bath, where $a$ can be related to the number of elements $N$ of the heat bath and tends to zero if $N$ is very large[18], which implies

$$f(x)=c[1+a\beta x]^{\frac{1}{a}} \xrightarrow[a\to 0]{} c e^{\beta x}.$$

This distribution is not a power law in the sense of Eq.(4). It has neither the scale invariance nor the translation invariance mentioned above. The operator on $x$ that keeps $f(x)$ invariant is a generalized addition $x+_a b = x+b+a\beta xb$ [19], i.e. $f(x+_a b)=c[1+a\beta(x+b+a\beta xb)]^{\frac{1}{a}}$ $=[1+a\beta b]^{\frac{1}{a}}c[1+a\beta x]^{\frac{1}{a}}=g(b)f(x)$ where $g(b)=[1+a\beta b]^{\frac{1}{a}}$.

The derivation of Eq.(5) from $f(x+_a b) = g(b)f(x)$ is given as follows. First we make a derivative such as $\frac{df(x+_a b)}{db} = \frac{df(x+_a b)}{d(x+_a b)}\frac{d(x+_a b)}{db} = \frac{df(x+_a b)}{d(x+_a b)}(1+a\beta x) = g'(b)f(x)$.

Then let $b=0$, we get $\frac{df(x)}{dx}(1+a\beta x) = g'(0)f(x)$ which means $\frac{df(x)}{f(x)} = g'(0)\frac{dx}{(1+a\beta x)}$ or $\ln f = \ln(1+a\beta x)^{1/a} + c$ or Eq.(5) with $\beta = g'(0)$.

### 3) A definition of entropy as a measure of dynamical uncertainty

Suppose we have a random (discrete) variable $x_i$ with a probability distribution $p_i = \frac{1}{Z} f(x_i)$ where $i$ is the state index. The average of $x_i$ is given by $\overline{x} = \sum_i x_i p_i$ and the normalization is $\sum_i p_i = 1$. The uncertainty in a probability distribution of $x$ can be measured by many quantities. The standard deviation $\sigma$ or the variance $\sigma^2 = \overline{x^2} - \overline{x}^2$, for example, can surely be used if they exist. Shannon entropy, as some of other known entropy forms, can also be used as a measure of uncertainty of any $p_i$. But certainly any given entropy form, including Shannon one, cannot be maximized for any distribution $p_i$ according to maxent rules, which can be seen below. The main task of this work, let us recall it, is to search for a general definition of uncertainty measure underlying maxent, in such a way that each derived entropy can be maximized to give the corresponding distribution. Here we propose a quite general measure $I$ by a variational definition as follows

$$dI = d\overline{x} - \overline{dx} = \sum_i x_i d p_i. \qquad (6)$$

where $\overline{dx} = \sum_i p_i dx_i$. This definition has been in a way inspired by the first and second laws of thermodynamics in equilibrium thermodynamics. Considering the definition of internal energy $\overline{E} = \sum_i p_i E_i$ where $E_i$ is the energy of the microstate $i$ with probability $p_i$, we can write

$\delta\overline{E} = \sum_i \delta p_i E_i + \sum_i p_i \delta E_i = \sum_i \delta p_i E_i + \overline{\delta E_i}$ in which $\overline{\delta E_i} = \sum_i p_i \delta E_i = \sum_j \left( \sum_i p_i \frac{\partial E_i}{\partial q_j} \right) \delta q_j$ is the work done to the system by external forces $F_j = \sum_i p_i \frac{\partial E_i}{\partial q_j}$ where $q_j$ is the extensive variables such as volume, distance or electrical polarization. According to the first law of thermodynamics, the quantity $\sum_i \delta p_i E_i = \delta\overline{E} - \overline{\delta E_i}$ is the heat change in the system, that is, $\sum_i \delta p_i E_i = \delta Q = T\delta S$ for a reversible process, where $S$ is the thermodynamic entropy and $T$ the absolute temperature. $S$ has the following variational relation

$$\delta S = \frac{1}{T}\left(\delta\overline{E} - \overline{\delta E}\right). \tag{7}$$

As is well known, $S$ measures the dynamical disorder of the system or the uncertainty of the probability distribution of energy in the dynamics. In contrast with another measure $\sigma^2 = \overline{E^2} - \overline{E}^2$, $S$ can be maximized in maxent for equilibrium system to derive probability distribution. At this point, we must mention a work by A. Plastino and EFM Curado [20] on the equivalence between the particular thermodynamic relation $\delta S = \beta\delta\overline{E}$ and maxent for probability assignment. In order to obtain this heat-energy relationship from the first and second laws of thermodynamics, they considered the particular reversible process affecting only the microstate population, i.e., the distribution $p_i$ has a variation $\delta p_i$ due to a heat transfer $\delta Q = T\delta S$. Although their variational approach from a particular process still needs further justification, their conclusion is important and consequential. Not only have they shown the equivalence between $\delta S = \beta\delta\overline{E}$ and maxent, they also reached a quite generic variational principle by virtue of the fundamental laws of thermodynamics. To our opinion, as a variational method, $\delta S = \beta\delta\overline{E}$ is much more general and powerful than maxent with originally a unique maximizable measure, since it enables one to maximize other form of entropy if any for equilibrium system, which opens the door for other development of statistical mechanics on the basis of maxent for equilibrium system.

Coming back to Eq.(6) which is just an extension of Eq.(7) to arbitrary random variables $x$ and to arbitrary system (even out of equilibrium) and a generic definition of maximizable uncertainty measure for any random variable. The maximizability of this measure consists to put $\overline{dx} = 0$ as discussed below.

The geometrical aspect of the uncertainty measure defined by Eq.(6) can be illustrated in the examples of Figure 1 which shows that $dI$ and $I$ are related to the width of the distributions on the one hand, and to the form of the distribution on the other. $dI$ is not an increasing function of the distribution width. For example, $dI=0$ for uniform distribution whatever the width of $p(x)$=constant. This means that $I$ is a constant.

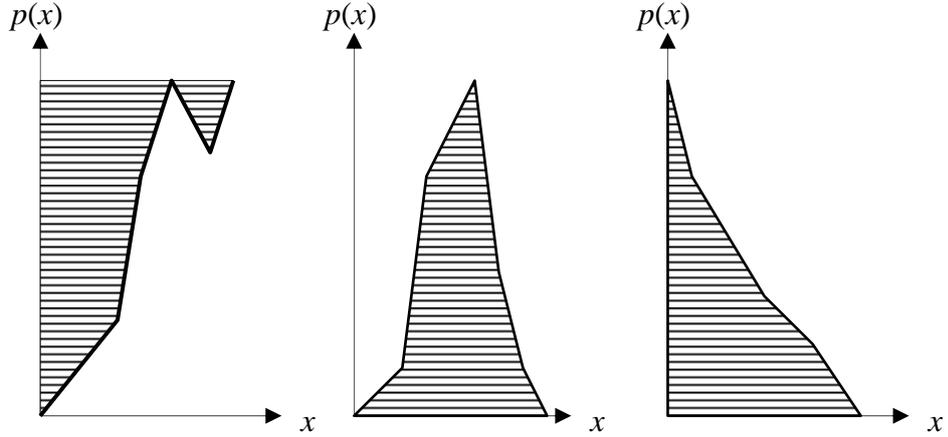

**Figure 1**, Three geometrical representations of the variation *dI* defined in Eq.(6) for some distributions. The hatched areas represent the value of the entropy variation $dI = d\bar{x} - \overline{dx} = \sum_i x_i dp_i$ for each case. In general, *dI* for any increasing or decreasing part of a distribution curve is the area between the *p(x)*-axis and the considered part. So in this figure, *dI* for any part of the distribution curve is the area from *p(x)*-axis to the left of the part. The sign of *dI* depends on the way it is calculated. If it is calculated with increasing *x*, *dI* is positive for the increasing part of *p(x)* since $dp_i > 0$ (main parts of left panel) and negative for the decreasing part of *p(x)* since $dp_i < 0$ (right panel). In the middle panel, the hatched area is the difference of $dI_{inc} > 0$ for the increasing part and the $dI_{dec} < 0$ for the decreasing part. Since the region of $dI_{inc}$ is smaller than that of $dI_{dec}$, the difference is negative. It is easily seen that *dI*=0 if the width (uncertainty) of the distribution is zero or if the distribution is uniform.

## 4) Three probability distribution laws and their entropy

In this section, on the basis of the uncertainty measure defined in Eq.(6), we derive the entropy functionals for the three probability laws discussed in section 2.

### a) Translation invariant probability and Shannon entropy

The following calculation is trivial. From Eq.(6), for exponential distribution $p_i = \frac{1}{Z} e^{-x_i}$, we have

$$dI = -\sum_i \ln(p_i Z) d\, p_i = -\sum_i \ln p_i d\, p_i - \ln Z \sum_i d\, p_i = -\sum_i \ln p_i d\, p_i = -d \sum_i p_i \ln p_i$$

and

$$I = -\sum_i p_i \ln p_i + c.$$

This is Shannon information if the constant $c$ is neglected. Within the conventional statistical mechanics, this is the Gibbs formula for Clausius entropy. Remember that the maximization of this entropy using Lagrange multiplier associated with expectation of $x$ yields exponential distribution law.

### b) Scale invariant probability and entropy functional

We have in this case power law probability distribution $p_i$:

$$p_i = \frac{1}{Z} x_i^{-a}. \tag{8}$$

where $Z = \sum_i x_i^{-a}$. Put it into Eq.(6) to get

$$dI = \sum_i (p_i Z)^{-1/a} d\, p_i = Z^{-1/a} \frac{1}{1-1/a} d\sum_i p_i^{1-1/a} = -Z^{-1/a} d[\sum_i p_i^{1-1/a}/(1-1/a) + c] \tag{9}$$

where $c$ is an arbitrary constant. Since we are addressing a given system to find its entropy form, $Z$ can be considered as a constant for the variation in $x$ (the reader will find below that this constant can be given by the Lagrange multiplier in the maximum entropy formalism). Hence we can write

$$I \propto \sum_i p_i^{1-1/a}/(1-1/a) + c. \tag{10}$$

In order to determine $c$, we imagine a system with two states $i=1$ and 2 with $p_1 = 0$ and $p_2 = 1$. In this case, $I=0$ so that

$$\left(\frac{0+1}{1-1/a}\right) + c = 0 \tag{11}$$

i.e.,

$$c = -\frac{1}{1-1/a} \tag{12}$$

We finally get

$$I = \frac{\sum_i p_i^{1-1/a} - 1}{1 - 1/a} \qquad (13)$$

Let $q = \frac{1}{a}$, we can write

$$I = -\frac{1 - \sum_i p_i^{1-q}}{1-q} = -\sum_i \frac{p_i - p_i^{1-q}}{1-q} \qquad (14)$$

Notice that this functional does not yield Shannon entropy for $q \to 1$. As a matter of fact, $q$ must be positive and smaller than unity. $I$ is negative if $q$ is greater than unity or smaller than zero, which does not make sense. For large $x$ Lévy flight for example, $1 < a < 3$, so $\frac{1}{3} < q < 1$.

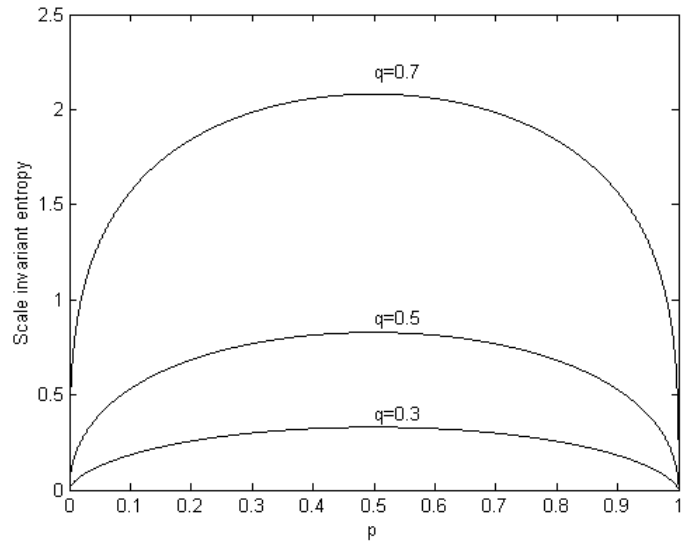

**Figure 2**, The variation of the scale invariant entropy $I = -\sum_{i=1}^{2} \frac{p_i - p_i^{1-q}}{1-q}$ with $p_1 = p$ and $p_2 = 1-p$ for different $q$ values. It can be shown that if $q \to 0$, $S = -\frac{q}{1-q}\sum p \ln p \to 0$, and If $q \to 1$, $S = \frac{\sum 1 - 1}{1-q} + \sum \ln p \to \infty$.

It can be calculated that

$$I = -\sum_i \frac{p_i - p_i^{1-q}}{1-q} = -\sum_i p_i \frac{1 - p_i^{-q}}{1-q} = -\sum_i p_i \frac{1 - \frac{1}{Z^{-q}}x}{1-q} = -\frac{1 - Z^q \bar{x}}{1-q} \quad (15)$$

Its behavior with probability is shown in Figure 2. The maximization of $I$ conditioned with a Lagrange multiplier $\beta$ such as $\delta(I - \beta \bar{x}) = 0$ directly yields the power law of Eq.(8) with $\beta = Z^q = Z^{1/a}$.

### c) The entropy for q-exponential probability

We have seen above that the probability $p_i = c[1 - a\beta x_i]^{\frac{1}{a}}$ had a special invariant property. Let us express $x$ as a function of $p_i$ and put it into Eq.(6) to get

$$dI = \sum_i \frac{1 - (p_i/c)^a}{a\beta} d p_i = -\frac{1}{a\beta c^a} \sum_i p_i^a d p_i = -\frac{1}{a\beta c^a (1+a)} d(\sum_i p_i^{1+a} + c) \quad (16)$$

By the same tricks for determining $c$ in the above section, we get $c=-1$. So we can write

$$I = -\frac{\sum_i p_i^{1+a} - 1}{a} = -\sum_i \frac{p_i - p_i^q}{1-q} \quad (17)$$

Where $q=1+a$ and we have used the normalization $\sum_i p_i = 1$. This is the Tsallis entropy which tends to the Shannon entropy for $q \to 1$ or $a \to 0$. In this case $p_i = c[1 - a\beta x_i]^{\frac{1}{a}}$ tends to an exponential distribution.

## 5) Concluding remarks

We have derived the entropy functionals for three probability distributions. This was done on the basis of a variational definition of uncertainty measure, or entropy without postulating entropy property (such as additivity) as in the usual information theory. The variational definition $dI = d\bar{x} - \overline{dx}$ is valid for any probability distributions of $x$ as long as it has finite expectation value. According to the results, the exponential probability has the Shannon entropy, the power law distribution has an entropy $I = -\frac{1 - \sum_i p_i^{1-q}}{1-q}$ where $0<q<1$, and the q-exponential distribution has Tsallis entropy $I = -\sum_i \frac{p_i - p_i^q}{1-q}$ where $q$ is positive.

It is worth mentioning again that the present definition of entropy as a measure of uncertainty offers the possibility of introducing the maximum entropy principle in a natural way with Lagrange multipliers associated with expectation of the random variables. It is easy to verify, with the above three entropies, that the maximum entropy calculus yields the original probability distributions. This is not an ordinary and fortuitous mutual invertibility, since the probability and the entropy are not reciprocal functions and the maximum entropy calculus is not a usual mathematical operation. As a matter of fact, this invertibility between entropy and probability resides in the variational definition $dI = d\bar{x} - \overline{dx}$. As discussed in the section 3, $\overline{dx}$ can be considered as an extended work whatever the nature of *x*. So to get the "equilibrium state" or stable probability distribution, we can put $\overline{dx}$=0 just as in the mechanical equilibrium condition where the vector sum of all forces acting on an object should be 0, an underlying idea of the principle of virtual work in mechanics. We straightforwardly get $dI-\beta d\bar{x}$=0 or $d(I+\alpha\sum_i p_i+\beta\bar{x})$=0 if we add the normalization condition.

This reasoning leads in a more natural way to the maxent idea $\delta S - \beta\delta\bar{E} = 0$ of Plastino and Curado[20] for thermodynamic equilibrium.

## Acknowledgements

We thank Professor F. Tsobnang and Professor S. Abe for useful discussion and comments. We would like to express here our thanks to the referees for their pertinent criticism and very useful comments and suggestions.